\documentclass[aps,prl,reprint,amsmath,amssymb,showpacs,superscriptaddress]{revtex4-2}
\usepackage{CJK}
\usepackage{graphicx}% Include figure files
\usepackage{dcolumn}% Align table columns on decimal point
\usepackage{bm}% bold math
\usepackage{color}
\usepackage{orcidlink}
\usepackage{hyperref}% add hypertext capabilities
\begin{document}
\begin{CJK*}{UTF8}{gbsn}
\title{Emergence of the halo in $^{11}$Li from full nuclear many-body dynamics
}% Force line breaks with \\
\author{Yilong Yang\orcidlink{0000-0002-5065-1309} ({\CJKfamily{gbsn}杨一龙})}
\affiliation{State Key Laboratory of Nuclear Physics and Technology, School of Physics, Peking University, Beijing 100871, China}

\author{Pengwei Zhao\orcidlink{0000-0001-8243-2381}  ({\CJKfamily{gbsn}赵鹏巍})}
\email{pwzhao@pku.edu.cn}
\affiliation{State Key Laboratory of Nuclear Physics and Technology, School of Physics, Peking University, Beijing 100871, China}

\begin{abstract}
The two-neutron halo nucleus $^{11}$Li is a paradigmatic quantum many-body system whose large spatial extent and weak binding have long challenged a microscopic description from first principles.  
Using a neural-network variational Monte Carlo approach, we present an \textit{ab initio} demonstration that the halo structure of $^{11}$Li emerges directly from the underlying nuclear interactions and full many-body dynamics. 
The calculation employs an essential nuclear Hamiltonian constrained solely by few-body observables and reproduces the binding and separation energies of Li isotopes, as well as the isotopic trend of their matter radii.
We identify a correlation between the halo size in $^{11}$Li and the splitting of $P$-wave neutron-alpha scattering phase shifts, establishing the crucial role of neutron-alpha spin-orbit interactions in halo formation. 
Dineutron correlations are found to arise naturally from the many-body wave function without assuming a preformed core-plus-valence-neutron structure. 
These results provide a microscopic understanding of halo formation in $^{11}$Li and establish a link between few-body scattering observables and emergent many-body structure.
\end{abstract}

\maketitle
\end{CJK*}
%*********************************************************%
%---------------------Introduction------------------------%
%*********************************************************%
\textit{Introduction}---Quantum halo systems, characterized by substantial components extending well into classically forbidden regions, are of particular interest in molecular, atomic, and nuclear physics~\cite{Jensen2004Rev.Mod.Phys.76.215261}. 
In nuclear physics, the two-neutron halo nucleus $^{11}$Li is one of the most celebrated examples. 
Experimentally, a sudden rise in the interaction cross sections was observed from $^{9}$Li to $^{11}$Li~\cite{Tanihata1985Phys.Rev.Lett.55.26762679}, indicating a matter radius far larger than expected from the conventional mass dependence $1.2A^{1/3}$. 
Subsequent measurements of charge radii, low-lying dipole strength, and knockout reactions further established its halo character~\cite{Sanchez2006Phys.Rev.Lett.96.033002,Nakamura2006Phys.Rev.Lett.96.252502,Kubota2020Phys.Rev.Lett.125.252501}.
Despite decades of studies, whether the halo in $^{11}$Li can emerge directly from the underlying nuclear interactions and full nuclear many-body dynamics remains unresolved. 
The challenge arises from the coexistence of strong many-body correlations, weak binding, and an extended density tail, resulting in strong coupling to the continuum.

Existing descriptions of $^{11}$Li are largely based on two complementary approximations to the nuclear many-body problem. 
Three-body models and halo effective field theory (EFT) describe the system as an inert $^{9}$Li core plus two valence neutrons~\cite{Hansen1995Ann.Rev.Nucl.Part.Sci.45.591634,Hammer2017J.Phys.G44.103002}, but do not fully account for core polarization and the Pauli-blocking effects. 
In contrast, (relativistic) mean-field approaches treat all nucleons as active degrees of freedom~\cite{Bertsch1989Phys.Rev.C.39.1154,Sagawa1992Phys.Lett.B286.712,Meng1996Phys.Rev.Lett.77.39633966,Meng2006Prog.Part.Nucl.Phys.57.470563}, but miss important correlations between the halo neutrons. 
Hybrid approaches combining mean-field and cluster descriptions have also been developed~\cite{Tomaselli2001Nucl.Phys.A690.298301,Varga2002Phys.Rev.C66.041302,Myo2008Prog.Theor.Phys.119.561581,Noertershaeuser2011Phys.Rev.C84.024307}. While these approaches successfully describe selected properties of $^{11}$Li, they do not address the emergence of the halo from the complete many-body dynamics.

A full $A$-body description of $^{11}$Li would provide a direct understanding of halo formation from the underlying nuclear interactions and many-body dynamics. 
The difficulty lies in the simultaneous description of strong short-range many-body correlations and the extended density tail associated with weak binding and continuum coupling. 
Methods based on harmonic-oscillator expansions, such as the no-core shell model (NCSM)~\cite{Barrett2013Prog.Part.Nucl.Phys.69.131181}, face challenges in describing the slowly decaying halo tail. 
Although several properties of $^{11}$Li have been reported within NCSM calculations~\cite{Navratil1998Phys.Rev.C57.31193128,Forssen2009Phys.Rev.C.79.021303,Caprio2022Phys.Rev.C105.L061302}, these calculations do not reproduce a positive and weakly bound two-neutron separation energy and have not addressed the halo structure directly.  
Coordinate-space approaches, including nuclear lattice EFT~\cite{Lee2009Prog.Part.Nucl.Phys.63.117154,Laedhe2019.} and continuum quantum Monte Carlo (QMC) methods~\cite{Carlson2015Rev.Mod.Phys.1067,Gandolfi2020Front.Phys.8.}, are in principle well suited for halo systems, but their applications to $^{11}$Li remain severely limited by the fermion sign problem. 
Significant progress has been achieved through NCSM combined with continuum approaches~\cite{RomeroRedondo2016Phys.Rev.Lett.117.222501,Calci2016Phys.Rev.Lett.117.242501,Navratil2016Phys.Scr.91.053002,navratil2026arXiv} and through nuclear lattice EFT with wave-function matching~\cite{Shen2025Phys.Rev.Lett.134.162503,Shen2026Particles9.,Elhatisari2024Nature630.5963}. 
Nevertheless, the emergence of the $^{11}$Li halo from full $A$-body dynamics has not yet been demonstrated.

Recently, deep-neural-network representations of nuclear many-body wave functions have significantly extended the capabilities of continuum QMC methods~\cite{Adams2021Phys.Rev.Lett.127.022502,lovato2026arXiv}. 
These approaches enable highly accurate variational solutions of light nuclei while avoiding the fermion sign problem that limits conventional diffusion QMC calculations~\cite{Lovato2022Phys.Rev.Research4.043178,Yang2023Phys.Rev.C107.034320}. 
They have successfully described a broad range of nuclear observables, including electromagnetic properties, response functions, relativistic effects, and scattering observables~\cite{Gnech2024Phys.Rev.Lett.133.142501,yang2025arXiv2509.01303,Parnes2026Phys.Rev.Lett.136.032501,Yang2022Phys.Lett.B835.137587,Yang2025ChinesePhys.Lett.42.051201,Yang2025Phys.Rev.Lett.135.172502}. 
However, their application to weakly bound halo nuclei has remained largely unexplored.

In this Letter, we use the deep-learning variational QMC method \textit{FeynmanNet} to demonstrate the emergence of the $^{11}$Li halo directly from the underlying nuclear interactions and full many-body dynamics.  
Employing an essential nuclear Hamiltonian constrained solely by few-body observables, we reproduce the binding energies, separation energies, and matter radii along the Li isotopic chain.
We show that the emergence of the halo is closely linked to the splitting of $P$-wave neutron-alpha scattering phase shifts, identifying neutron-alpha spin-orbit interactions as a key ingredient of halo formation. 
In addition, dineutron correlations are found to arise naturally from the many-body wave function without assuming a preformed core-plus-valence-neutron structure.

\textit{FeynmanNet}---The deep-neural-network wave function takes the Slater-Jastrow-backflow form~\cite{Yang2023Phys.Rev.C107.034320},
\begin{equation}
    \Psi(\bm X)=e^{\mathcal{U}(\bm X)}\sum_{n=1}^{N_{\rm det}}\det \left[\mathbf f^{(n)}(\bm X)\odot\mathbf \Phi^{(n)}(\bm X)\right],
\end{equation}
where $\bm X=(\bm x_1,\ldots,\bm x_A)$ with $\bm x_i=(\bar{\bm r}_i,s_i,t_i)$ are the single-nucleon spatial, spin, and isospin coordinates, $\mathcal{U}$ is a neural-network Jastrow factor, and $N_{\rm det}$ is the number of backflow-transformed Slater determinants.
The intrinsic coordinates $\bar{\bm r}_i=\bm r_i-\bm r_{\rm cm}$ ensure exact translational invariance of the wave function.
The matrices $\mathbf f$ and $\mathbf \Phi$ ($\odot$ denotes element-wise multiplication) denote backflow-transformed orbitals and single-nucleon shell-model orbitals, respectively.
Embedding the major shell structure through $\mathbf \Phi$ significantly improves the efficiency of the neural-network wave function and accelerates the training process.

The backflow-transformed orbitals depend on both single- and pair-nucleon latent variables, allowing many-body correlations to be encoded directly in the determinants. 
The construction of the orbitals preserves permutation symmetry, such that exchanging two nucleons is equivalent to swapping two rows of the determinant matrix, thereby guaranteeing the antisymmetry of the wave function.
Further details of the architecture can be found in Ref.~\cite{Yang2023Phys.Rev.C107.034320}.

Different from Ref.~\cite{Yang2023Phys.Rev.C107.034320}, where separate neural networks are used for the real and imaginary parts of the many-body wave function, we instead represent the real and imaginary components of the backflow-transformed orbitals by separate neural networks,
$\mathbf f^{(n)}=\mathbf f^{({\rm R},n)}+i\mathbf f^{({\rm I},n)}$. 
This halves the number of determinants required in the wave-function representation.

The neural-network wave function is optimized within variational Monte Carlo using the stochastic reconfiguration method~\cite{Sorella2005Phys.Rev.B71.241103}. 
The robustness of the variational minima has been verified through different initial wave functions and three statistically independent training runs.
For $^{11}$Li, initial states with matter radii differing by approximately 3 fm converge to final radii consistent within 0.03 fm. 
The number of determinants is taken to be $N_{\rm det}=4$ for $^{6\text{--}9}$Li and $N_{\rm det}=8$ for $^{11}$Li. 
The residual uncertainties associated with $N_{\rm det}$ are estimated to be below 0.1 MeV for ground-state energies and 0.02 fm for matter radii. 
Additional details on the neural-network architecture, optimization procedure, and convergence tests are provided in the Supplemental Material~\cite{Supp}.

\textit{Nuclear Hamiltonian}---The Hamiltonian contains two-nucleon (NN) and three-nucleon (3N) interactions,
\begin{equation}\label{eq.H}
    \begin{split}
        H&=\sum_{i=1}^A\frac{-\nabla^2_i}{2M}+\sum_{i<j}(v^{\rm EM}_{ij}+v^{\rm c}_{ij}+v^{\rm so}_{ij})+\sum_{i<j<k}V^{\rm c}_{ijk}.
    \end{split}
\end{equation}
Here, $v^{\rm EM}$ denotes the Coulomb interaction between finite-size protons~\cite{Wiringa1995Phys.Rev.C51.3851}. 
The central NN interaction $v^{\rm c}$ is model ``o'' of Ref.~\cite{Schiavilla2021Phys.Rev.C103.054003}, inspired by the leading-order pionless EFT expansion. 
It contains four parameters that are fixed to the $np$ scattering lengths and effective ranges in the $S/T=0/1$ and $1/0$ channels.

The central 3N interaction $V^{\rm c}_{ijk}$ contains only a short-range term with two parameters, its coupling strength $c_E$ and regulator $R_3$.
In Refs.~\cite{Schiavilla2021Phys.Rev.C103.054003, Gnech2024Phys.Rev.Lett.133.142501},
the $c_E$ value was fixed by the binding energy of $^3$H at a given $R_3$, and $R_3$ was then adjusted to the binding energies of $^{16}$O and heavier nuclei.
In contrast, here we constrain the 3N interaction only by $A\le 4$ observables, determining its coupling strength $c_E=1.5514$ and regulator $R_3=1.2$~fm exclusively from the binding energies of $^3$H and $^4$He.
Since $^3$H and $^4$He do not constrain the $T=3/2$ component of the 3N interaction, which is not determined within the present low-energy calibration, we retain only its $T=1/2$ component by applying an isospin projection operator.

The resulting Hamiltonian containing only the central NN and 3N interactions is referred to hereafter as the \emph{essential Hamiltonian}, emphasizing that it is a minimal low-energy Hamiltonian constrained by few-body observables rather than a complete order-by-order EFT interaction.
However, as indicated in the previous study~\cite{Gnech2024Phys.Rev.Lett.133.142501}, such a Hamiltonian systematically underbinds open-shell $p$-shell nuclei, indicating the need for additional spin-dependent interactions.

To remedy this deficiency, we extend the essential Hamiltonian by including a short-range spin-orbit interaction,
\begin{equation}\label{eq.vso}
    v^{\rm so}_{ij}=-C_{\rm so}\frac{\partial_r \delta(\bm r_{ij})}{r_{ij}}\bm L\cdot\bm S,
\end{equation}
where $\bm r_{ij}=\bm r_i-\bm r_j$, and $\bm L$ and $\bm S$ are the relative angular momentum and total spin of the nucleon pair, respectively. 
The $\delta$ function is regularized using the same regulator as the central NN interaction in the $S/T=1/0$ channel~\cite{Schiavilla2021Phys.Rev.C103.054003}. 
The spin-orbit term does not contribute to $S$-wave NN scattering and therefore preserves the successful description of low-energy NN observables provided by the central interaction.
The Hamiltonian including the spin-orbit interaction is referred to hereafter as the \emph{improved essential Hamiltonian}.

The spin-orbit strength $C_{\rm so}$ is fixed exclusively by the splitting of low-energy $P$-wave neutron-alpha scattering phase shifts, without using any information from Li isotopes.
Following Ref.~\cite{Yang2025Phys.Rev.Lett.135.172502}, the phase shifts are calculated by placing the five nucleons in a harmonic-oscillator trap, converting the scattering problem into an eigenvalue problem solvable with variational Monte Carlo. 
The phase shifts are then extracted using the Busch-Englert-Rza\.{z}ewski-Wilkens formula~\cite{Busch1998FoundationsofPhysics28.549559,Suzuki2009Phys.Rev.A80.033601}.

\begin{figure}[!htbp]
    \centering
    \includegraphics[width=0.9\linewidth]{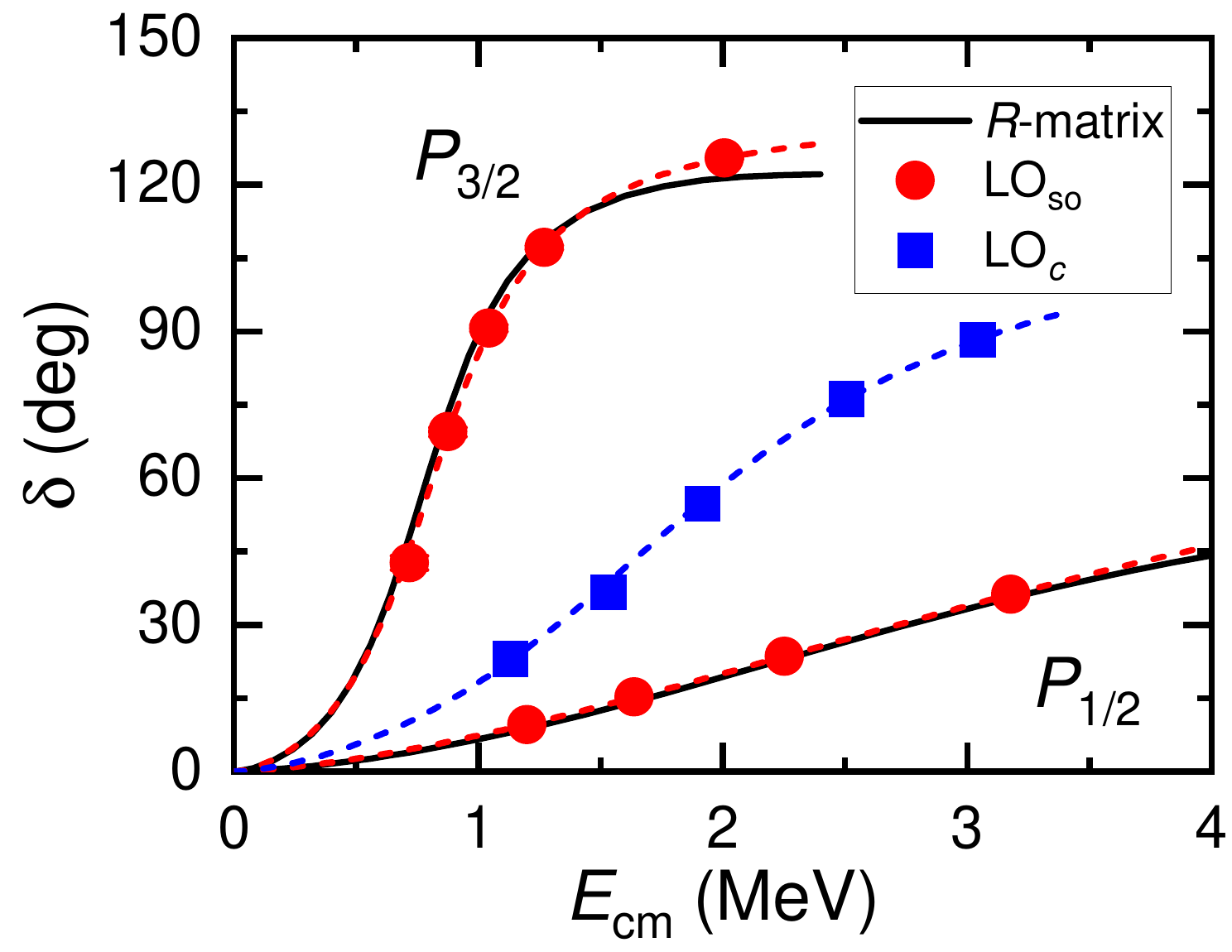}
   \caption{$P$-wave neutron-alpha scattering phase shifts as functions of center-of-mass scattering energy $E_{\rm cm}$, compared with the $R$-matrix analysis of experimental data~\cite{Hale2025}. LO$_{\rm c}$ and LO$_{\rm so}$ denote the results from the essential Hamiltonian and improved essential Hamiltonian, respectively. The spin-orbit strength $C_{\rm so}$ is determined from the splitting between the $P_{3/2}$ and $P_{1/2}$ phase shifts.}
    \label{fig1}
\end{figure}

Figure~\ref{fig1} compares the calculated low-energy $P$-wave neutron-alpha scattering phase shifts with the $R$-matrix analysis of experimental data~\cite{Hale2025}. 
The central Hamiltonian alone yields degenerate $P_{3/2}$ and $P_{1/2}$ phase shifts, in contrast to the substantial experimental splitting. 
Such splitting arises from spin-dependent components of nuclear interactions beyond leading central terms~\cite{Lynn2016Phys.Rev.Lett.116.062501}. 
In the improved essential Hamiltonian, these effects are incorporated effectively through the short-range spin-orbit interaction. 
A single parameter, $C_{\rm so}=-2.8~{\rm fm}^4$, reproduces both the low-energy $P_{3/2}$ and $P_{1/2}$ phase shifts. 
As shown below, the same spin-orbit interaction that reproduces neutron-alpha scattering also governs the emergence of the halo in $^{11}$Li.

%\textit{Emergence of Halo in $^{11}$Li}---Figure~\ref{fig2}(a) depicts the neutron separation energies along the Li isotopic chain predicted by \textit{FeynmanNet} using the improved essential Hamiltonian.
%The theoretical uncertainties are estimated by three statistically independent runs of the training procedure and are smaller than symbols.
%The present results reproduce well the experimental data, with root-mean-square (rms) deviation $\sigma=0.36(2)$ MeV.
%The total binding energies are also in excellent agreement, with rms deviation $\sigma=0.23(1)$ MeV.
%In addition, the present calculations do not yield a bound $^{10}$Li, consistent with experiments.

\begin{figure}[!htbp]
    \centering
    \includegraphics[width=0.9\linewidth]{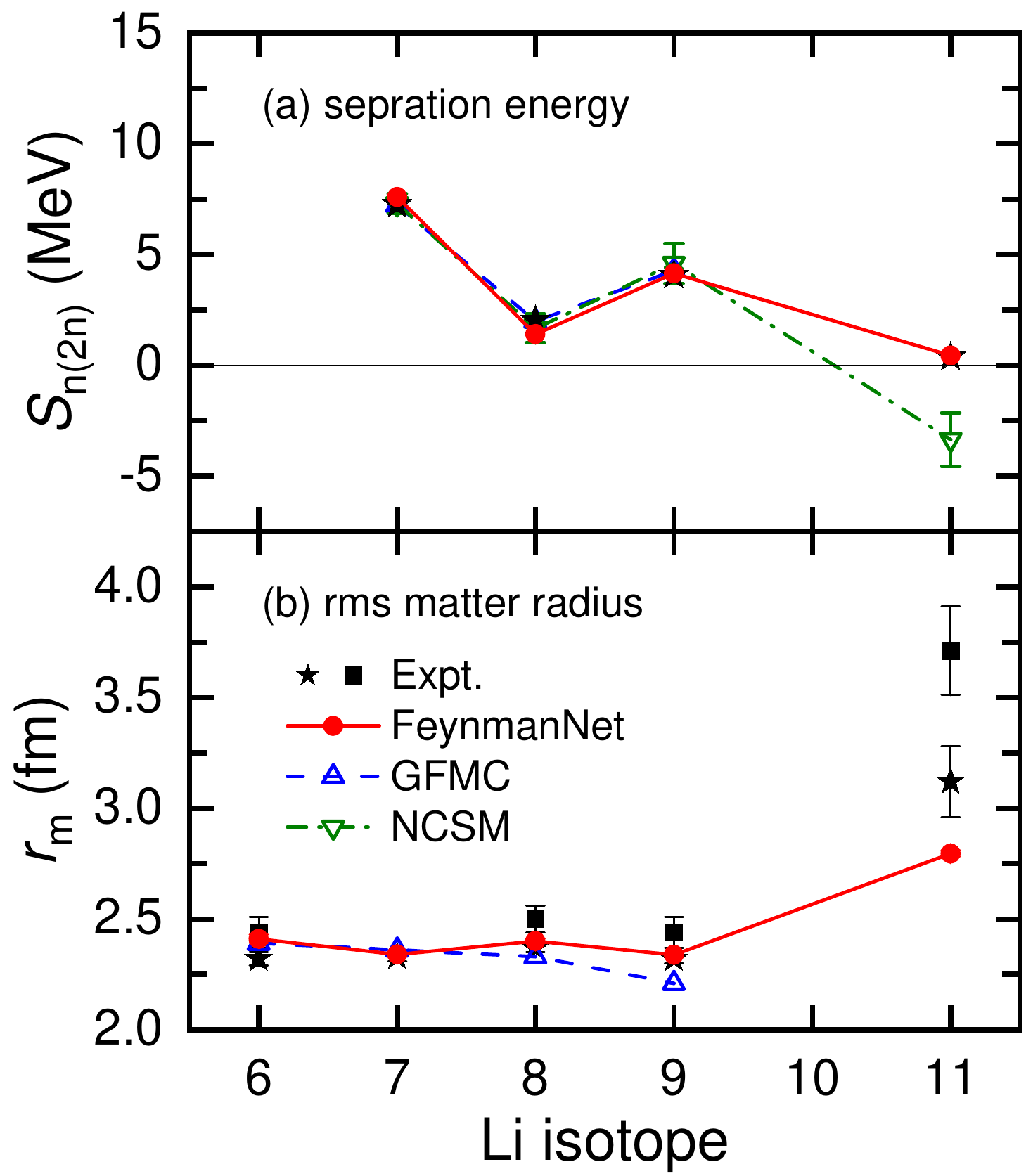}   
    \caption{(a) Neutron separation energies $S_n$ (two-neutron separation energy $S_{2n}$ for $^{11}$Li) predicted by \textit{FeynmanNet} using the improved essential Hamiltonian, compared with experimental data~\cite{Wang2021Chin.Phys.C45.030003} and existing no-core shell model and Green's function Monte Carlo calculations~\cite{Carlson2015Rev.Mod.Phys.1067,Forssen2009Phys.Rev.C.79.021303}. The calculations reproduce the weak binding of the halo nucleus $^{11}$Li. (b) Matter radii along the Li isotopic chain. The sharp increase from $^{9}$Li to $^{11}$Li signals the emergence of the halo. Experimental values are extracted from interaction cross sections (star)~\cite{Tanihata1988Phys.Lett.B206.592596} and proton scattering data (square)~\cite{Dobrovolsky2006Nucl.Phys.A766.124}, and their error bars denote mainly statistical errors.
}
    \label{fig2}
\end{figure}

\textit{Emergence of Halo in $^{11}$Li}---The improved essential Hamiltonian reproduces the neutron separation energies along the Li isotopic chain, including the weak binding of the halo nucleus $^{11}$Li, as seen in Fig.~\ref{fig2}(a). 
The theoretical uncertainties, estimated from three statistically independent training runs, are smaller than the symbol size. 
The neutron separation energies agree well with experiment, with a root-mean-square (rms) deviation of $\sigma=0.36(2)$ MeV, and the total binding energies are reproduced with comparable accuracy. 
In addition, the present calculations do not yield a bound $^{10}$Li, consistent with experiment.

The results are compared with existing \textit{ab initio} calculations using Green's function Monte Carlo (GFMC)~\cite{Carlson2015Rev.Mod.Phys.1067} and the no-core shell model (NCSM)~\cite{Forssen2009Phys.Rev.C.79.021303}. 
While all approaches describe the lighter Li isotopes reasonably well, only the present calculations reproduce a small but positive two-neutron separation energy for $^{11}$Li. 
Existing GFMC calculations have not yet been extended to $^{11}$Li, whereas available NCSM calculations predict the nucleus to be unbound.

Figure~\ref{fig2}(b) depicts the matter radii along the Li isotopic chain. 
The calculated radii remain nearly constant from $^{6}$Li to $^{9}$Li and then increase sharply in $^{11}$Li, signaling the emergence of a halo.
Quantitatively, the calculated radii agree with experiment within uncertainties for $^{6\text{--}9}$Li but remain somewhat smaller than the reported values for $^{11}$Li.
The remaining discrepancy may reflect the absence of explicit pion-exchange interactions in the present Hamiltonian, which could enhance the spatial extent of the halo. 
It may also originate from the model dependence involved in extracting matter radii from reaction observables, particularly for nuclei with extended density tails~\cite{Tanihata1988Phys.Lett.B206.592596,Dobrovolsky2006Nucl.Phys.A766.124}.

\begin{figure}[!htbp]
    \centering
    \includegraphics[width=0.95\linewidth]{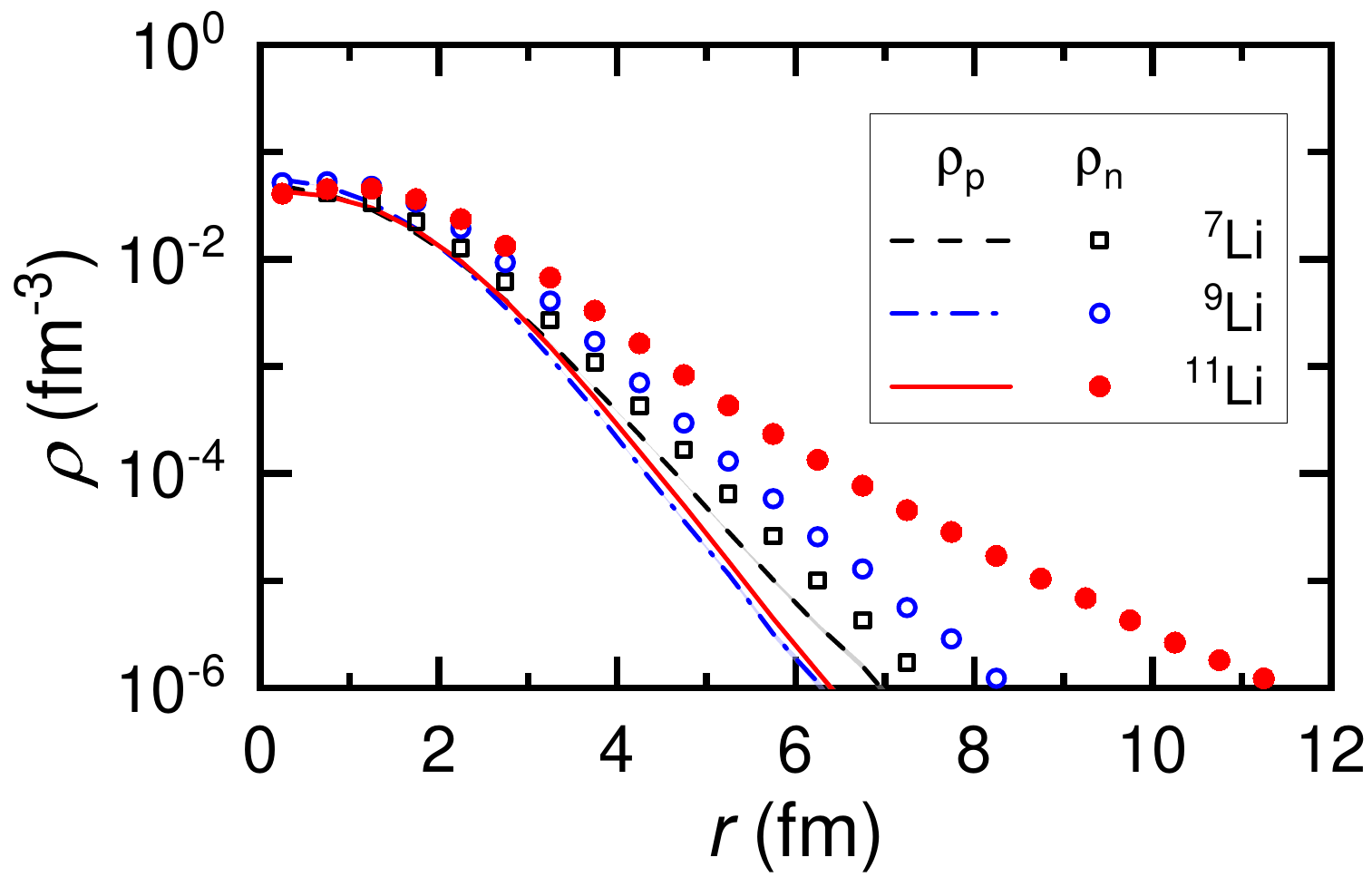}
    \caption{Proton (lines) and neutron (dots) density distributions of $^7$Li, $^9$Li, and $^{11}$Li calculated with the improved essential Hamiltonian.}
    \label{fig3}
\end{figure}

The origin of the enhanced matter radius is illustrated in Fig.~\ref{fig3}, which shows the proton and neutron density distributions for $^{7}$Li, $^{9}$Li, and $^{11}$Li. 
As neutrons are added along the isotopic chain, the neutron distribution evolves from a compact profile in $^{7}$Li to a neutron skin in $^{9}$Li and finally develops a slowly decaying tail in $^{11}$Li. 
The emergence of this extended neutron density tail directly demonstrates the development of the halo structure in $^{11}$Li.

\begin{figure*}[!htbp]
    \centering
    \includegraphics[width=0.8\linewidth]{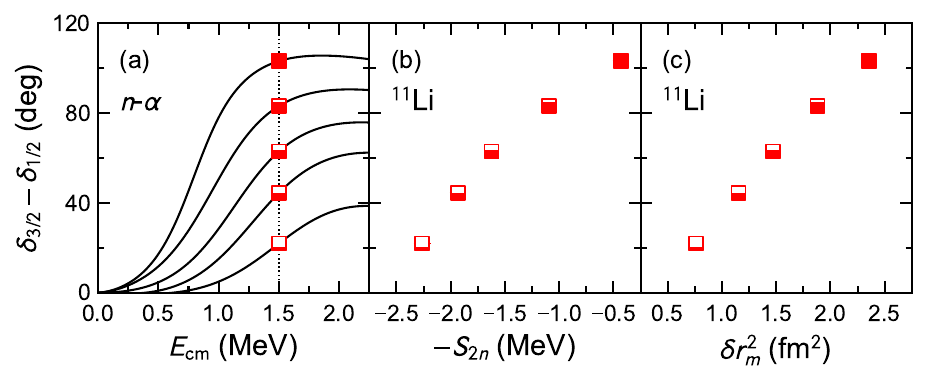}
    \caption{(a) Splitting between the computed $P_{3/2}$ ($\delta_{3/2}$) and $P_{1/2}$ ($\delta_{1/2}$) neutron-alpha scattering phase 
    shifts for different spin-orbit coupling strengths. 
    The symbol filling indicates the spin-orbit strength, with increasing fill corresponding to increasing interaction strength; 
    the fully filled symbol corresponds to $C_{\rm so}=-2.8~\mathrm{fm}^4$, which reproduces the experimental phase-shift splitting.
    (b) Correlation between the two-neutron separation energy $S_{2n}$ of $^{11}$Li and the phase-shift splitting $\delta_{3/2}-\delta_{1/2}$ at $E_{\rm cm}=1.5$~MeV. (c) Same as (b), but for the mean-square matter-radius difference $\delta r_m^2=(r_m^{11})^2-(r_m^9)^2$.
    The same symbol filling is used across all panels to identify calculations with the same spin-orbit coupling strength.}
    \label{fig4}
\end{figure*}

To identify the microscopic origin of the halo, we investigate how the halo properties of $^{11}$Li evolve with the neutron-alpha spin-orbit interaction.
A clear connection emerges between the halo observables and the splitting of the $P$-wave neutron-alpha scattering phase shifts, which serves as a measure of the spin-orbit component of the neutron-alpha interaction.

Figure~\ref{fig4} shows results obtained by varying the spin-orbit strength over the range $C_{\rm so}=-0.7$ to $-2.8$ fm$^4$, where the latter value is the strength that reproduces the neutron-alpha scattering data (Fig.~\ref{fig1}). 
As the spin-orbit interaction increases, the splitting between the $P_{3/2}$ and $P_{1/2}$ phase shifts grows [Fig.~\ref{fig4}(a)]. 
Simultaneously, the two-neutron separation energy of $^{11}$Li approaches zero [Fig.~\ref{fig4}(b)], while the mean-square matter-radius difference $\delta r_m^2=(r_m^{11})^2-(r_m^9)^2$ increases rapidly [Fig.~\ref{fig4}(c)]. 
The same spin-orbit interaction that reproduces neutron-alpha scattering therefore drives the weak binding and large spatial extent characteristic of the halo nucleus $^{11}$Li.
These results demonstrate a direct correlation between the size of the halo in $^{11}$Li and the splitting of free-space $P$-wave neutron-alpha scattering phase shifts, and reveal the crucial role of neutron-alpha spin-orbit interactions in halo formation.

The observed correlation can be understood qualitatively from the underlying single-particle structure. 
Within the conventional single-particle picture, the two halo neutrons predominantly occupy the $1p_{1/2}$ orbital. 
As the neutron-alpha spin-orbit interaction becomes stronger, the splitting between the $1p_{1/2}$ and $1p_{3/2}$ orbitals increases, driving the $1p_{1/2}$ orbital closer to threshold. 
The resulting weaker binding leads to a more extended single-particle wave function and promotes the formation of the halo.

\textit{Dineutron correlations}---Unlike three-body descriptions~\cite{Esbensen1997Phys.Rev.C56.30543062,Hagino2005Phys.Rev.C.72.044321,Hagino2007Phys.Rev.Lett.99.022506}, the present many-body wave function does not assume a preformed $^9$Li core and two halo neutrons. 
To investigate whether dineutron correlations emerge from the full $A$-body dynamics, we analyze the opening-angle distribution of the two halo neutrons,
\begin{equation}\label{eq.costheta}
    \cos\theta_{ nn}=\frac{\bm r_{c1}\cdot\bm r_{c2}}{|\bm r_{c1}||\bm r_{c2}|},
\end{equation}
where $\bm r_{cn}=\bm r_n-\bm r_c$ ($n=1,2$) is the position of a halo neutron relative to the center of mass of the $^9$Li core. 
For uncorrelated neutrons, the probability distribution is proportional to $\frac12\sin\theta_{nn}$ and peaks at $\theta_{nn}=90^\circ$, whereas dineutron correlations enhance the probability at small opening angles. 

Halo neutrons are identified directly from Monte Carlo samples of the full many-body wave function. 
Configurations are generated from $|\Psi(\bm X)|^2$ using the Metropolis-Hastings algorithm~\cite{Metropolis1953TheJournalofChemicalPhysics21.10871092,Hastings1970Biometrika57.97109}. 
For each configuration, the three protons and the six neutrons closest to their center of mass are assigned to the $^9$Li core, while the two most distant neutrons are identified as halo neutrons. 
The opening angle $\theta_{nn}$ is then evaluated event by event and accumulated into a probability distribution. 
A total of $10^6$ configurations are analyzed.

\begin{figure}[!htbp]
    \centering
    \includegraphics[width=0.95\linewidth]{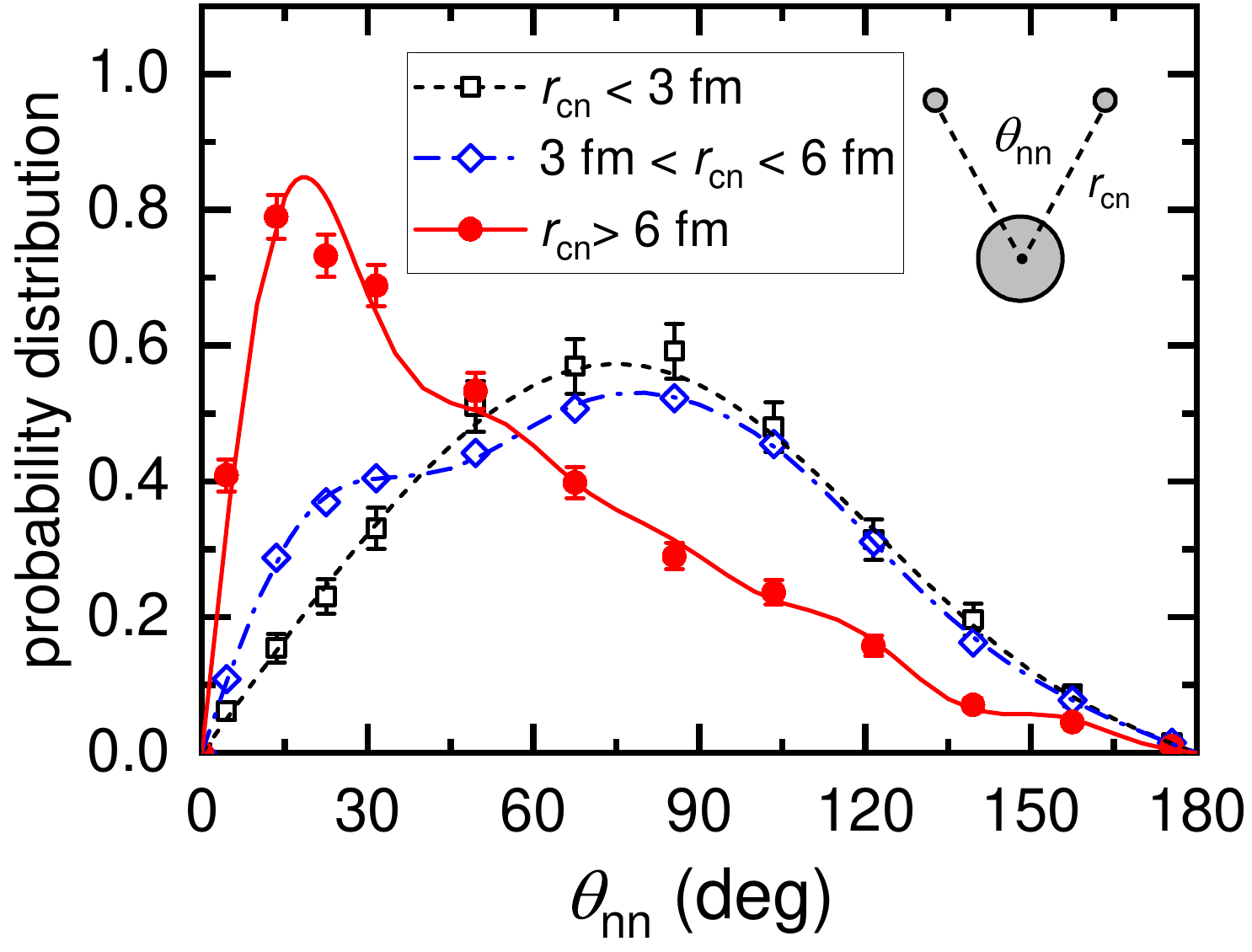}
     \caption{Probability distributions of the opening angle $\theta_{nn}$ for the two halo neutrons in $^{11}$Li. Results are shown for three regions defined by the neutron-core distance $r_{cn}$.
     The black empty squares, blue empty diamonds, and red solid dots correspond to $r_{cn}<3\ {\rm fm}$, $3\ {\rm fm}<r_{cn}<6\ {\rm fm}$, and $r_{cn}>6\ {\rm fm}$, respectively. Solid lines are Fourier sine series fits to the Monte Carlo results to guide the eye.
          }
    \label{fig5}
\end{figure}

Figure~\ref{fig5} depicts the opening-angle distributions for the two halo neutrons in three regions defined by their distance from the core. 
A clear evolution is observed as the halo neutrons move away from the core. 
For $r_{cn}<3$ fm, the distribution peaks near $\theta_{nn}=90^\circ$ and closely resembles that of uncorrelated neutrons. 
A shoulder first develops at small opening angles in the intermediate region $3<r_{cn}<6$ fm, indicating the onset of dineutron correlations. 
In the peripheral region $r_{cn}>6$ fm, a pronounced peak emerges near $\theta_{nn}=0^\circ$, demonstrating strong dineutron correlations in the halo.

This evolution can be understood from the competition between neutron-core and neutron-neutron correlations. 
Near the core, the halo neutrons overlap significantly with the $^9$Li core, and neutron-core interactions together with the Pauli principle dominate their dynamics. 
As the neutrons move into the low-density halo region, these effects become progressively weaker, allowing neutron-neutron correlations to emerge and eventually dominate.

In Ref.~\cite{Kubota2020Phys.Rev.Lett.125.252501}, the $(p,pn)$ knockout measurement showed that the mean correlation angle between the two valence neutrons peaks at a missing momentum $k\sim 0.3$~fm$^{-1}$ and decreases at both smaller and larger $k$. 
Within a three-body reaction model~\cite{Kikuchi2016Prog.Theor.Exp.Phys.103D03}, this behavior was interpreted as a dineutron localized near the nuclear surface $r\sim 3.6$~fm, rather than in the core interior or the halo tail.
A similar surface localization was found in an earlier three-body structure calculation~\cite{Hagino2007Phys.Rev.Lett.99.022506}.
Our full $A$-body calculations, however, show that the dineutron correlation continues to strengthen in the halo tail. 
This difference highlights that inferring the spatial distribution of dineutron correlations from a reaction observable depends on both the reaction dynamics and the structural assumptions entering its interpretation.
A direct comparison with experiment therefore requires reaction calculations based on the same full $A$-body wave functions~\cite{Horiuchi2026Phys.Rev.Lett.136.202501,Horiuchi2026Phys.Rev.C113.064601}.

\textit{Summary}---Using the deep-learning variational quantum Monte Carlo method \textit{FeynmanNet}, we have demonstrated that the halo structure of $^{11}$Li emerges directly from the underlying nuclear interactions and full $A$-body dynamics.
Employing an improved essential Hamiltonian constrained solely by few-body observables, we obtain a realistic description of the Li isotopic chain, including the weak binding and enhanced matter radius of the halo nucleus $^{11}$Li.
The calculations reveal a direct correlation between the size of the halo and the splitting of free-space $P$-wave neutron-alpha scattering phase shifts, identifying neutron-alpha spin-orbit interactions as a key ingredient of halo formation. 
In addition, dineutron correlations are found to emerge naturally from the full $A$-body wave function without assuming a preformed core-plus-valence-neutron structure.
With the favorable polynomial scaling of \textit{FeynmanNet}, the present framework can be extended to other weakly bound nuclei and medium-mass neutron-rich systems. 
More generally, the present work demonstrates a quantitative link between a free-space scattering observable and the emergence of halo structure in a many-body system, providing a microscopic bridge between few-body physics and emergent many-body structure.

%*********************************************************%
%-----------------------Summary---------------------------%
%*********************************************************%
% Acknowledgement
\begin{acknowledgments}
We thank J{\o}rgen Randrup and Yiping Wang for helpful discussions and suggestions.
This work has been supported in part by the National Key R\&D Program of China (Contracts No. 2024YFA1612600, No. 2024YFE0109803), 
the National Natural Science Foundation of China (Grants No. 12475117 and No. 12435006), the National Key Laboratory of Neutron Science and Technology (Grant No. NST202401016), and the High-performance Computing Platform of Peking University.
\end{acknowledgments}

\bibliography{refs_Li}

@article{Kikuchi2016Prog.Theor.Exp.Phys.103D03,
    author = {Kikuchi, Yuma and Ogata, Kazuyuki and Kubota, Yuki and Sasano, Masaki and Uesaka, Tomohiro},
    title = {{Determination of a dineutron correlation in Borromean nuclei via a quasi-free knockout (p,pn) reaction}},
    journal = {Prog. Theor. Exp. Phys.},
    volume = {2016},
    number = {10},
    pages = {103D03},
    year = {2016},
    month = {10},
    issn = {2050-3911},
    doi = {10.1093/ptep/ptw148},
    url = {https://doi.org/10.1093/ptep/ptw148},
    eprint = {https://academic.oup.com/ptep/article-pdf/2016/10/103D03/9621191/ptw148.pdf},
}

@article{Horiuchi2026Phys.Rev.C113.064601,
  title = {{Glauber-theory analysis of nuclear reactions on a $^{12}\mathrm{C}$ target with variational Monte Carlo wave functions}},
  author = {Horiuchi, W. and Suzuki, Y. and Wiringa, R. B.},
  journal = {Phys. Rev. C},
  volume = {113},
  issue = {6},
  pages = {064601},
  numpages = {21},
  year = {2026},
  month = {Jun},
  publisher = {American Physical Society},
  doi = {10.1103/gcbk-s7tc},
  url = {https://link.aps.org/doi/10.1103/gcbk-s7tc}
}

@article{Horiuchi2026Phys.Rev.Lett.136.202501,
  title = {{Glauber-Theory Calculations of High-Energy Nuclear Scattering Observables Using Variational Monte Carlo Wave Functions}},
  author = {Horiuchi, W. and Suzuki, Y. and Wiringa, R. B.},
  journal = {Phys. Rev. Lett.},
  volume = {136},
  issue = {20},
  pages = {202501},
  numpages = {6},
  year = {2026},
  month = {May},
  publisher = {American Physical Society},
  doi = {10.1103/ppqx-yn59},
  url = {https://link.aps.org/doi/10.1103/ppqx-yn59}
}

@Article{Wiringa1995Phys.Rev.C51.3851,
  author    = {Wiringa, R. B. and Stoks, V. G. J. and Schiavilla, R.},
  journal   = {Phys. Rev. C},
  title     = {{Accurate nucleon-nucleon potential with charge-independence breaking}},
  year      = {1995},
  month     = {Jan},
  pages     = {38--51},
  volume    = {51},
  doi       = {10.1103/PhysRevC.51.38},
  issue     = {1},
  numpages  = {0},
  publisher = {American Physical Society},
}

@Article{Schiavilla2021Phys.Rev.C103.054003,
  author    = {Schiavilla, R. and Girlanda, L. and Gnech, A. and Kievsky, A. and Lovato, A. and Marcucci, L. E. and Piarulli, M. and Viviani, M.},
  journal   = {Phys. Rev. C},
  title     = {{Two- and three-nucleon contact interactions and ground-state energies of light- and medium-mass nuclei}},
  year      = {2021},
  month     = {May},
  pages     = {054003},
  volume    = {103},
  doi       = {10.1103/PhysRevC.103.054003},
  issue     = {5},
  numpages  = {18},
  publisher = {American Physical Society},
  url       = {https://link.aps.org/doi/10.1103/PhysRevC.103.054003},
}

@Article{Wang2021Chin.Phys.C45.030003,
  author    = {Wang, Meng and Huang, W.J. and Kondev, F.G. and Audi, G. and Naimi, S.},
  journal   = {Chin. Phys. C},
  title     = {{The AME 2020 atomic mass evaluation (II). Tables, graphs and references*}},
  year      = {2021},
  month     = {mar},
  number    = {3},
  pages     = {030003},
  volume    = {45},
  doi       = {10.1088/1674-1137/abddaf},
  publisher = {Chinese Physical Society and the Institute of High Energy Physics of the Chinese Academy of Sciences and the Institute of Modern Physics of the Chinese Academy of Sciences and IOP Publishing Ltd},
  url       = {https://doi.org/10.1088/1674-1137/abddaf},
}

@Article{Navratil1998Phys.Rev.C57.31193128,
  author    = {Navr\'atil, P. and Barrett, B. R.},
  journal   = {Phys. Rev. C},
  title     = {{Large-basis shell-model calculations for $p$-shell nuclei}},
  year      = {1998},
  month     = {Jun},
  pages     = {3119--3128},
  volume    = {57},
  doi       = {10.1103/PhysRevC.57.3119},
  issue     = {6},
  numpages  = {0},
  publisher = {American Physical Society},
  url       = {https://link.aps.org/doi/10.1103/PhysRevC.57.3119},
}

@Article{Caprio2022Phys.Rev.C105.L061302,
  author    = {Caprio, Mark A. and Fasano, Patrick J. and Maris, Pieter},
  journal   = {Phys. Rev. C},
  title     = {{Robust ab initio prediction of nuclear electric quadrupole observables by scaling to the charge radius}},
  year      = {2022},
  month     = {Jun},
  pages     = {L061302},
  volume    = {105},
  doi       = {10.1103/PhysRevC.105.L061302},
  issue     = {6},
  numpages  = {7},
  publisher = {American Physical Society},
  url       = {https://link.aps.org/doi/10.1103/PhysRevC.105.L061302},
}

@Article{Navratil2016Phys.Scr.91.053002,
  author    = {Petr Navr\'atil and Sofia Quaglioni and Guillaume Hupin and Carolina Romero-Redondo and Angelo Calci},
  journal   = {Phys. Scr.},
  title     = {{Unified ab initio approaches to nuclear structure and reactions}},
  year      = {2016},
  month     = {apr},
  number    = {5},
  pages     = {053002},
  volume    = {91},
  doi       = {10.1088/0031-8949/91/5/053002},
  publisher = {IOP Publishing},
  url       = {https://dx.doi.org/10.1088/0031-8949/91/5/053002},
}

@misc{navratil2026arXiv,
      title={{Halo Nuclei from Ab Initio Nuclear Theory}}, 
      author={Petr Navratil and Sofia Quaglioni and Guillaume Hupin and Michael Gennari and Kostas Kravvaris},
      year={2026},
      eprint={2604.02612},
      archivePrefix={arXiv},
      primaryClass={nucl-th},
      url={https://arxiv.org/abs/2604.02612}, 
}

@Article{Noertershaeuser2011Phys.Rev.C84.024307,
  author    = {N\"ortersh\"auser, W. and Neff, T. and S\'anchez, R. and Sick, I.},
  journal   = {Phys. Rev. C},
  title     = {{Charge radii and ground state structure of lithium isotopes: Experiment and theory reexamined}},
  year      = {2011},
  month     = {Aug},
  pages     = {024307},
  volume    = {84},
  doi       = {10.1103/PhysRevC.84.024307},
  issue     = {2},
  numpages  = {14},
  publisher = {American Physical Society},
  url       = {https://link.aps.org/doi/10.1103/PhysRevC.84.024307},
}

@article{Calci2016Phys.Rev.Lett.117.242501,
  title = {{Can Ab Initio Theory Explain the Phenomenon of Parity Inversion in $^{11}\mathrm{Be}$?}},
  author = {Calci, Angelo and Navr\'atil, Petr and Roth, Robert and Dohet-Eraly, J\'er\'emy and Quaglioni, Sofia and Hupin, Guillaume},
  journal = {Phys. Rev. Lett.},
  volume = {117},
  issue = {24},
  pages = {242501},
  numpages = {6},
  year = {2016},
  month = {Dec},
  publisher = {American Physical Society},
  doi = {10.1103/PhysRevLett.117.242501},
  url = {https://link.aps.org/doi/10.1103/PhysRevLett.117.242501}
}

@Article{Lee2009Prog.Part.Nucl.Phys.63.117154,
  author   = {Dean Lee},
  journal  = {Prog. Part. Nucl. Phys.},
  title    = {{Lattice simulations for few- and many-body systems}},
  year     = {2009},
  issn     = {0146-6410},
  number   = {1},
  pages    = {117-154},
  volume   = {63},
  doi      = {https://doi.org/10.1016/j.ppnp.2008.12.001},
  url      = {https://www.sciencedirect.com/science/article/pii/S014664100800094X},
}

@Book{Laedhe2019.,
  author    = {L\"adhe, Timo A. and Mei\ss{}ner, Ulf-G.},
  publisher = {Springer Cham},
  title     = {{Nuclear Lattice Effective Field Theory: An Introduction}},
  year      = {2019},
  series    = {Lect. Note Phys.},
  doi       = {https://doi.org/10.1007/978-3-030-14189-9},
}

@Article{Gandolfi2020Front.Phys.8.,
  author  = {Gandolfi, Stefano and Lonardoni, Diego and Lovato, Alessandro and Piarulli, Maria},
  journal = {Front. Phys.},
  title   = {{Atomic Nuclei From Quantum Monte Carlo Calculations With Chiral EFT Interactions}},
  year    = {2020},
  issn    = {2296-424X},
  volume  = {8},
  doi     = {10.3389/fphy.2020.00117},
  url     = {https://www.frontiersin.org/articles/10.3389/fphy.2020.00117},
}

@Article{Elhatisari2024Nature630.5963,
  author   = {Elhatisari, Serdar and Bovermann, Lukas and Ma, Yuan-Zhuo and Epelbaum, Evgeny and Frame, Dillon and Hildenbrand, Fabian and Kim, Myungkuk and Kim, Youngman and Krebs, Hermann and L\"ahde, Timo A. and Lee, Dean and Li, Ning and Lu, Bing-Nan and Mei\ss{}ner, Ulf-G. and Rupak, Gautam and Shen, Shihang and Song, Young-Ho and Stellin, Gianluca},
  journal  = {Nature},
  title    = {{Wavefunction matching for solving quantum many-body problems}},
  year     = {2024},
  issn     = {1476-4687},
  number   = {8015},
  pages    = {59--63},
  volume   = {630},
  doi      = {10.1038/s41586-024-07422-z},
  refid    = {Elhatisari2024},
  url      = {https://doi.org/10.1038/s41586-024-07422-z},
}

@Article{Carlson2015Rev.Mod.Phys.1067,
  author    = {Carlson, J. and Gandolfi, S. and Pederiva, F. and Pieper, Steven C. and Schiavilla, R. and Schmidt, K. E. and Wiringa, R. B.},
  journal   = {Rev. Mod. Phys.},
  title     = {{Quantum Monte Carlo methods for nuclear physics}},
  year      = {2015},
  month     = {Sep},
  pages     = {1067--1118},
  volume    = {87},
  doi       = {10.1103/RevModPhys.87.1067},
  issue     = {3},
  numpages  = {52},
  publisher = {American Physical Society},
}

@Article{Yang2023Phys.Rev.C107.034320,
  author    = {Yang, Y. L. and Zhao, P. W.},
  journal   = {Phys. Rev. C},
  title     = {{Deep-neural-network approach to solving the ab initio nuclear structure problem}},
  year      = {2023},
  month     = {Mar},
  pages     = {034320},
  volume    = {107},
  doi       = {10.1103/PhysRevC.107.034320},
  issue     = {3},
  numpages  = {9},
  publisher = {American Physical Society},
  url       = {https://link.aps.org/doi/10.1103/PhysRevC.107.034320},
}

@Article{Lovato2022Phys.Rev.Research4.043178,
  author    = {Lovato, Alessandro and Adams, Corey and Carleo, Giuseppe and Rocco, Noemi},
  journal   = {Phys. Rev. Research},
  title     = {{Hidden-nucleons neural-network quantum states for the nuclear many-body problem}},
  year      = {2022},
  month     = {Dec},
  pages     = {043178},
  volume    = {4},
  doi       = {10.1103/PhysRevResearch.4.043178},
  issue     = {4},
  numpages  = {8},
  publisher = {American Physical Society},
  url       = {https://link.aps.org/doi/10.1103/PhysRevResearch.4.043178},
}

@Article{Sorella2005Phys.Rev.B71.241103,
  author    = {Sorella, Sandro},
  journal   = {Phys. Rev. B},
  title     = {{Wave function optimization in the variational Monte Carlo method}},
  year      = {2005},
  month     = {Jun},
  pages     = {241103},
  volume    = {71},
  doi       = {10.1103/PhysRevB.71.241103},
  issue     = {24},
  numpages  = {4},
  publisher = {American Physical Society},
}

@Article{Jensen2004Rev.Mod.Phys.76.215261,
  author    = {Jensen, A. S. and Riisager, K. and Fedorov, D. V. and Garrido, E.},
  journal   = {Rev. Mod. Phys.},
  title     = {{Structure and reactions of quantum halos}},
  year      = {2004},
  month     = {Feb},
  pages     = {215--261},
  volume    = {76},
  doi       = {10.1103/RevModPhys.76.215},
  issue     = {1},
  numpages  = {0},
  publisher = {American Physical Society},
  url       = {https://link.aps.org/doi/10.1103/RevModPhys.76.215},
}

@Article{Hansen1995Ann.Rev.Nucl.Part.Sci.45.591634,
  author    = {Hansen, P G and Jensen, A S and Jonson, B},
  journal   = {Ann. Rev. Nucl. Part. Sci.},
  title     = {{Nuclear Halos}},
  year      = {1995},
  issn      = {1545-4134},
  number    = {Volume 45,},
  pages     = {591-634},
  volume    = {45},
  doi       = {https://doi.org/10.1146/annurev.ns.45.120195.003111},
  publisher = {Annual Reviews},
  type      = {Journal Article},
  url       = {https://www.annualreviews.org/content/journals/10.1146/annurev.ns.45.120195.003111},
}

@Article{Meng2006Prog.Part.Nucl.Phys.57.470563,
  author   = {J. Meng and H. Toki and S. G. Zhou and S. Q. Zhang and W. H. Long and L. S. Geng},
  journal  = {Prog. Part. Nucl. Phys.},
  title    = {{Relativistic continuum Hartree Bogoliubov theory for ground-state properties of exotic nuclei}},
  year     = {2006},
  issn     = {0146-6410},
  number   = {2},
  pages    = {470 - 563},
  volume   = {57},
  doi      = {https://doi.org/10.1016/j.ppnp.2005.06.001},
}

@Article{Tanihata1985Phys.Rev.Lett.55.26762679,
  author    = {Tanihata, I. and Hamagaki, H. and Hashimoto, O. and Shida, Y. and Yoshikawa, N. and Sugimoto, K. and Yamakawa, O. and Kobayashi, T. and Takahashi, N.},
  journal   = {Phys. Rev. Lett.},
  title     = {{Measurements of Interaction Cross Sections and Nuclear Radii in the Light $p$-Shell Region}},
  year      = {1985},
  month     = {Dec},
  pages     = {2676--2679},
  volume    = {55},
  doi       = {10.1103/PhysRevLett.55.2676},
  issue     = {24},
  numpages  = {0},
  publisher = {American Physical Society},
  url       = {https://link.aps.org/doi/10.1103/PhysRevLett.55.2676},
}

@Article{Meng1996Phys.Rev.Lett.77.39633966,
  author    = {Meng, J. and Ring, P.},
  journal   = {Phys. Rev. Lett.},
  title     = {{Relativistic Hartree-Bogoliubov Description of the Neutron Halo in ${}^{11}$Li}},
  year      = {1996},
  month     = {Nov},
  pages     = {3963--3966},
  volume    = {77},
  doi       = {10.1103/PhysRevLett.77.3963},
  issue     = {19},
  numpages  = {0},
  publisher = {American Physical Society},
  url       = {https://link.aps.org/doi/10.1103/PhysRevLett.77.3963},
}

@Article{Gnech2024Phys.Rev.Lett.133.142501,
  author    = {Gnech, Alex and Fore, Bryce and Tropiano, Anthony J. and Lovato, Alessandro},
  journal   = {Phys. Rev. Lett.},
  title     = {{Distilling the Essential Elements of Nuclear Binding via Neural-Network Quantum States}},
  year      = {2024},
  month     = {Oct},
  pages     = {142501},
  volume    = {133},
  doi       = {10.1103/PhysRevLett.133.142501},
  issue     = {14},
  numpages  = {7},
  publisher = {American Physical Society},
  url       = {https://link.aps.org/doi/10.1103/PhysRevLett.133.142501},
}

@Article{Yang2022Phys.Lett.B835.137587,
  author   = {Y.L. Yang and P.W. Zhao},
  journal  = {Phys. Lett. B},
  title    = {{A consistent description of the relativistic effects and three-body interactions in atomic nuclei}},
  year     = {2022},
  issn     = {0370-2693},
  pages    = {137587},
  volume   = {835},
  doi      = {10.1016/j.physletb.2022.137587},
  url      = {https://www.sciencedirect.com/science/article/pii/S0370269322007213},
}

@Article{Yang2025ChinesePhys.Lett.42.051201,
  author    = {Yang, Yi-Long and Zhao, Peng-Wei},
  journal   = {Chinese Phys. Lett.},
  title     = {{Reconciling Light Nuclei and Nuclear Matter: Relativistic ab initio Calculations}},
  year      = {2025},
  month     = {apr},
  number    = {5},
  pages     = {051201},
  volume    = {42},
  doi       = {10.1088/0256-307X/42/5/051201},
  publisher = {Chinese Physical Society and IOP Publishing Ltd},
  url       = {https://dx.doi.org/10.1088/0256-307X/42/5/051201},
}

@misc{lovato2026arXiv,
      title={{Neural-network quantum states for the nuclear many-body problem}}, 
      author={Alessandro Lovato and Giuseppe Carleo and Bryce Fore and Morten Hjorth-Jensen and Jane Kim and Arnau Rios and Noemi Rocco},
      year={2026},
      eprint={2602.13826},
      archivePrefix={arXiv},
      primaryClass={nucl-th},
      url={https://arxiv.org/abs/2602.13826}, 
}

@misc{yang2025arXiv2509.01303,
      title={{Zemach radii and nuclear structure effects in hyperfine splitting of Lithium}}, 
      author={Yilong Yang and Evgeny Epelbaum and Chen Ji and Pengwei Zhao},
      year={2025},
      eprint={2509.01303},
      archivePrefix={arXiv},
      primaryClass={nucl-th},
      url={https://arxiv.org/abs/2509.01303}, 
}

@Article{Parnes2026Phys.Rev.Lett.136.032501,
  author    = {Parnes, Elad and Barnea, Nir and Carleo, Giuseppe and Lovato, Alessandro and Rocco, Noemi and Zhang, Xilin},
  journal   = {Phys. Rev. Lett.},
  title     = {{Nuclear Responses with Neural-Network Quantum States}},
  year      = {2026},
  month     = {Jan},
  pages     = {032501},
  volume    = {136},
  doi       = {10.1103/tlqz-nw28},
  issue     = {3},
  numpages  = {9},
  publisher = {American Physical Society},
  url       = {https://link.aps.org/doi/10.1103/tlqz-nw28},
}

@Article{Yang2025Phys.Rev.Lett.135.172502,
  author    = {Yang, Yilong and Epelbaum, Evgeny and Meng, Jie and Meng, Lu and Zhao, Pengwei},
  journal   = {Phys. Rev. Lett.},
  title     = {{Chiral Symmetry and Peripheral Neutron-$\ensuremath{\alpha}$ Scattering}},
  year      = {2025},
  month     = {Oct},
  pages     = {172502},
  volume    = {135},
  doi       = {10.1103/45g7-bmp6},
  issue     = {17},
  numpages  = {7},
  publisher = {American Physical Society},
  url       = {https://link.aps.org/doi/10.1103/45g7-bmp6},
}

@Article{Barrett2013Prog.Part.Nucl.Phys.69.131181,
  author   = {Bruce R. Barrett and Petr Navr\'atil and James P. Vary},
  journal  = {Prog. Part. Nucl. Phys.},
  title    = {{Ab initio no core shell model}},
  year     = {2013},
  issn     = {0146-6410},
  pages    = {131-181},
  volume   = {69},
  doi      = {https://doi.org/10.1016/j.ppnp.2012.10.003},
  url      = {https://www.sciencedirect.com/science/article/pii/S0146641012001184},
}

@Article{Hammer2017J.Phys.G44.103002,
  author    = {Hammer, H-W and Ji, C and Phillips, D R},
  journal   = {J. Phys. G},
  title     = {{Effective field theory description of halo nuclei}},
  year      = {2017},
  month     = {sep},
  number    = {10},
  pages     = {103002},
  volume    = {44},
  doi       = {10.1088/1361-6471/aa83db},
  publisher = {IOP Publishing},
  url       = {https://doi.org/10.1088/1361-6471/aa83db},
}

@Article{Adams2021Phys.Rev.Lett.127.022502,
  author    = {Adams, Corey and Carleo, Giuseppe and Lovato, Alessandro and Rocco, Noemi},
  journal   = {Phys. Rev. Lett.},
  title     = {{Variational Monte Carlo Calculations of $A\ensuremath{\le}4$ Nuclei with an Artificial Neural-Network Correlator Ansatz}},
  year      = {2021},
  month     = {Jul},
  pages     = {022502},
  volume    = {127},
  doi       = {10.1103/PhysRevLett.127.022502},
  issue     = {2},
  numpages  = {6},
  publisher = {American Physical Society},
  url       = {https://link.aps.org/doi/10.1103/PhysRevLett.127.022502},
}

@article{Forssen2009Phys.Rev.C.79.021303,
  title = {{Charge radii and electromagnetic moments of Li and Be isotopes from the ab initio no-core shell model}},
  author = {Forss\'en, C. and Caurier, E. and Navr\'atil, P.},
  journal = {Phys. Rev. C},
  volume = {79},
  issue = {2},
  pages = {021303},
  numpages = {5},
  year = {2009},
  month = {Feb},
  publisher = {American Physical Society},
  doi = {10.1103/PhysRevC.79.021303},
  url = {https://link.aps.org/doi/10.1103/PhysRevC.79.021303}
}

@misc{Hale2025,
    author = {Hale, G. M.},
    note = {private communication (2025).}
}

@Article{Dobrovolsky2006Nucl.Phys.A766.124,
  author   = {A.V. Dobrovolsky and G.D. Alkhazov and M.N. Andronenko and A. Bauchet and P. Egelhof and S. Fritz and H. Geissel and C. Gross and A.V. Khanzadeev and G.A. Korolev and G. Kraus and A.A. Lobodenko and G. M\"unzenberg and M. Mutterer and S.R. Neumaier and T. Sch\"afer and C. Scheidenberger and D.M. Seliverstov and N.A. Timofeev and A.A. Vorobyov and V.I. Yatsoura},
  journal  = {Nucl. Phys. A},
  title    = {{Study of the nuclear matter distribution in neutron-rich Li isotopes}},
  year     = {2006},
  issn     = {0375-9474},
  pages    = {1-24},
  volume   = {766},
  doi      = {https://doi.org/10.1016/j.nuclphysa.2005.11.016},
  url      = {https://www.sciencedirect.com/science/article/pii/S0375947405012017},
}

@Article{Tanihata1988Phys.Lett.B206.592596,
  author   = {I. Tanihata and T. Kobayashi and O. Yamakawa and S. Shimoura and K. Ekuni and K. Sugimoto and N. Takahashi and T. Shimoda and H. Sato},
  journal  = {Phys. Lett. B},
  title    = {{Measurement of interaction cross sections using isotope beams of Be and B and isospin dependence of the nuclear radii}},
  year     = {1988},
  issn     = {0370-2693},
  number   = {4},
  pages    = {592-596},
  volume   = {206},
  doi      = {https://doi.org/10.1016/0370-2693(88)90702-2},
  url      = {https://www.sciencedirect.com/science/article/pii/0370269388907022},
}

@Article{Varga2002Phys.Rev.C66.041302,
  author    = {Varga, K. and Suzuki, Y. and Lovas, R. G.},
  journal   = {Phys. Rev. C},
  title     = {{Microscopic multicluster model of ${}^{9,10,11}\mathrm{Li}$}},
  year      = {2002},
  month     = {Oct},
  pages     = {041302},
  volume    = {66},
  doi       = {10.1103/PhysRevC.66.041302},
  issue     = {4},
  numpages  = {4},
  publisher = {American Physical Society},
  url       = {https://link.aps.org/doi/10.1103/PhysRevC.66.041302},
}

@Article{Tomaselli2001Nucl.Phys.A690.298301,
  author  = {M. Tomaselli and S. Fritzsche and A. Dax and P. Egelhof and C. Kozhuharov and T. K\"uhl and D. Marx and M. Mutterer and S.R. Neumaier and W. N\"ortersh\"auser and H. Wang and H.-J. Kluge},
  journal = {Nucl. Phys. A},
  title   = {{Microscopic model for charge and matter distributions of nuclei}},
  year    = {2001},
  issn    = {0375-9474},
  note    = {Proc. Int. Symp. on Nuclei and Nucleons},
  number  = {1},
  pages   = {298-301},
  volume  = {690},
  doi     = {https://doi.org/10.1016/S0375-9474(01)00963-0},
  url     = {https://www.sciencedirect.com/science/article/pii/S0375947401009630},
}

@Article{Myo2008Prog.Theor.Phys.119.561581,
  author   = {Myo, Takayuki and Kikuchi, Yuma and Kat\={o}, Kiyoshi and Toki, Hiroshi and Ikeda, Kiyomi},
  journal  = {Prog. Theor. Phys.},
  title    = {{Systematic Study of 9,10,11Li with the Tensor and Pairing Correlations}},
  year     = {2008},
  issn     = {0033-068X},
  month    = {04},
  number   = {4},
  pages    = {561-581},
  volume   = {119},
  doi      = {10.1143/PTP.119.561},
  eprint   = {https://academic.oup.com/ptp/article-pdf/119/4/561/5156966/119-4-561.pdf},
  url      = {https://doi.org/10.1143/PTP.119.561},
}

@Article{Metropolis1953TheJournalofChemicalPhysics21.10871092,
  author  = {Metropolis,Nicholas and Rosenbluth,Arianna W. and Rosenbluth,Marshall N. and Teller,Augusta H. and Teller,Edward},
  journal = {The Journal of Chemical Physics},
  title   = {{Equation of State Calculations by Fast Computing Machines}},
  year    = {1953},
  number  = {6},
  pages   = {1087-1092},
  volume  = {21},
  doi     = {10.1063/1.1699114},
  eprint  = {https://doi.org/10.1063/1.1699114},
  url     = {https://doi.org/10.1063/1.1699114},
}

@Article{Hastings1970Biometrika57.97109,
  author   = {Hastings, W. K.},
  journal  = {Biometrika},
  title    = {{Monte Carlo sampling methods using Markov chains and their applications}},
  year     = {1970},
  issn     = {0006-3444},
  month    = {04},
  number   = {1},
  pages    = {97-109},
  volume   = {57},
  doi      = {10.1093/biomet/57.1.97},
  eprint   = {https://academic.oup.com/biomet/article-pdf/57/1/97/23940249/57-1-97.pdf},
  url      = {https://doi.org/10.1093/biomet/57.1.97},
}

@Article{Kubota2020Phys.Rev.Lett.125.252501,
  author    = {Kubota, Y. and Corsi, A. and Authelet, G. and Baba, H. and Caesar, C. and Calvet, D. and Delbart, A. and Dozono, M. and Feng, J. and Flavigny, F. and Gheller, J.-M. and Gibelin, J. and Giganon, A. and Gillibert, A. and Hasegawa, K. and Isobe, T. and Kanaya, Y. and Kawakami, S. and Kim, D. and Kikuchi, Y. and Kiyokawa, Y. and Kobayashi, M. and Kobayashi, N. and Kobayashi, T. and Kondo, Y. and Korkulu, Z. and Koyama, S. and Lapoux, V. and Maeda, Y. and Marqu\'es, F. M. and Motobayashi, T. and Miyazaki, T. and Nakamura, T. and Nakatsuka, N. and Nishio, Y. and Obertelli, A. and Ogata, K. and Ohkura, A. and Orr, N. A. and Ota, S. and Otsu, H. and Ozaki, T. and Panin, V. and Paschalis, S. and Pollacco, E. C. and Reichert, S. and Rouss\'e, J.-Y. and Saito, A. T. and Sakaguchi, S. and Sako, M. and Santamaria, C. and Sasano, M. and Sato, H. and Shikata, M. and Shimizu, Y. and Shindo, Y. and Stuhl, L. and Sumikama, T. and Sun, Y. L. and Tabata, M. and Togano, Y. and Tsubota, J. and Yang, Z. H. and Yasuda, J. and Yoneda, K. and Zenihiro, J. and Uesaka, T.},
  journal   = {Phys. Rev. Lett.},
  title     = {{Surface Localization of the Dineutron in $^{11}\mathrm{Li}$}},
  year      = {2020},
  month     = {Dec},
  pages     = {252501},
  volume    = {125},
  doi       = {10.1103/PhysRevLett.125.252501},
  issue     = {25},
  numpages  = {7},
  publisher = {American Physical Society},
  url       = {https://link.aps.org/doi/10.1103/PhysRevLett.125.252501},
}

@Article{Nakamura2006Phys.Rev.Lett.96.252502,
  author    = {Nakamura, T. and Vinodkumar, A. M. and Sugimoto, T. and Aoi, N. and Baba, H. and Bazin, D. and Fukuda, N. and Gomi, T. and Hasegawa, H. and Imai, N. and Ishihara, M. and Kobayashi, T. and Kondo, Y. and Kubo, T. and Miura, M. and Motobayashi, T. and Otsu, H. and Saito, A. and Sakurai, H. and Shimoura, S. and Watanabe, K. and Watanabe, Y. X. and Yakushiji, T. and Yanagisawa, Y. and Yoneda, K.},
  journal   = {Phys. Rev. Lett.},
  title     = {{Observation of Strong Low-Lying $E1$ Strength in the Two-Neutron Halo Nucleus $^{11}\mathrm{Li}$}},
  year      = {2006},
  month     = {Jun},
  pages     = {252502},
  volume    = {96},
  doi       = {10.1103/PhysRevLett.96.252502},
  issue     = {25},
  numpages  = {4},
  publisher = {American Physical Society},
  url       = {https://link.aps.org/doi/10.1103/PhysRevLett.96.252502},
}

@Article{Sanchez2006Phys.Rev.Lett.96.033002,
  author    = {S\'anchez, R. and N\"ortersh\"auser, W. and Ewald, G. and Albers, D. and Behr, J. and Bricault, P. and Bushaw, B. A. and Dax, A. and Dilling, J. and Dombsky, M. and Drake, G. W. F. and G\"otte, S. and Kirchner, R. and Kluge, H.-J. and K\"uhl, Th. and Lassen, J. and Levy, C. D. P. and Pearson, M. R. and Prime, E. J. and Ryjkov, V. and Wojtaszek, A. and Yan, Z.-C. and Zimmermann, C.},
  journal   = {Phys. Rev. Lett.},
  title     = {{Nuclear Charge Radii of $^{9,11}\mathrm{Li}$: The Influence of Halo Neutrons}},
  year      = {2006},
  month     = {Jan},
  pages     = {033002},
  volume    = {96},
  doi       = {10.1103/PhysRevLett.96.033002},
  issue     = {3},
  numpages  = {4},
  publisher = {American Physical Society},
  url       = {https://link.aps.org/doi/10.1103/PhysRevLett.96.033002},
}

@Article{RomeroRedondo2016Phys.Rev.Lett.117.222501,
  author    = {Romero-Redondo, Carolina and Quaglioni, Sofia and Navr\'atil, Petr and Hupin, Guillaume},
  journal   = {Phys. Rev. Lett.},
  title     = {{How Many-Body Correlations and $\ensuremath{\alpha}$ Clustering Shape $^{6}\mathrm{He}$}},
  year      = {2016},
  month     = {Nov},
  pages     = {222501},
  volume    = {117},
  doi       = {10.1103/PhysRevLett.117.222501},
  issue     = {22},
  numpages  = {5},
  publisher = {American Physical Society},
  url       = {https://link.aps.org/doi/10.1103/PhysRevLett.117.222501},
}

@Article{Shen2025Phys.Rev.Lett.134.162503,
  author    = {Shen, Shihang and Elhatisari, Serdar and Lee, Dean and Mei\ss{}ner, Ulf-G. and Ren, Zhengxue},
  journal   = {Phys. Rev. Lett.},
  title     = {{Ab Initio Study of the Beryllium Isotopes $^{7}\mathrm{Be}$ to $^{12}\mathrm{Be}$}},
  year      = {2025},
  month     = {Apr},
  pages     = {162503},
  volume    = {134},
  doi       = {10.1103/PhysRevLett.134.162503},
  issue     = {16},
  numpages  = {7},
  publisher = {American Physical Society},
  url       = {https://link.aps.org/doi/10.1103/PhysRevLett.134.162503},
}

@Article{Busch1998FoundationsofPhysics28.549559,
  author   = {Busch, Thomas and Englert, Berthold-Georg and Rza\.zewski, Kazimierz and Wilkens, Martin},
  journal  = {Foundations of Physics},
  title    = {{Two Cold Atoms in a Harmonic Trap}},
  year     = {1998},
  issn     = {1572-9516},
  number   = {4},
  pages    = {549--559},
  volume   = {28},
  doi      = {10.1023/A:1018705520999},
  refid    = {Busch1998},
  url      = {https://doi.org/10.1023/A:1018705520999},
}

@Article{Suzuki2009Phys.Rev.A80.033601,
  author    = {Suzuki, Akira and Liang, Yi and Bhaduri, Rajat K.},
  journal   = {Phys. Rev. A},
  title     = {{Two-atom energy spectrum in a harmonic trap near a Feshbach resonance at higher partial waves}},
  year      = {2009},
  month     = {Sep},
  pages     = {033601},
  volume    = {80},
  doi       = {10.1103/PhysRevA.80.033601},
  issue     = {3},
  numpages  = {6},
  publisher = {American Physical Society},
  url       = {https://link.aps.org/doi/10.1103/PhysRevA.80.033601},
}

@Article{Lynn2016Phys.Rev.Lett.116.062501,
  author    = {Lynn, J. E. and Tews, I. and Carlson, J. and Gandolfi, S. and Gezerlis, A. and Schmidt, K. E. and Schwenk, A.},
  journal   = {Phys. Rev. Lett.},
  title     = {{Chiral Three-Nucleon Interactions in Light Nuclei, Neutron-$\ensuremath{\alpha}$ Scattering, and Neutron Matter}},
  year      = {2016},
  month     = {Feb},
  pages     = {062501},
  volume    = {116},
  doi       = {10.1103/PhysRevLett.116.062501},
  issue     = {6},
  numpages  = {5},
  publisher = {American Physical Society},
  url       = {https://link.aps.org/doi/10.1103/PhysRevLett.116.062501},
}

@Article{Hagino2007Phys.Rev.Lett.99.022506,
  author    = {Hagino, K. and Sagawa, H. and Carbonell, J. and Schuck, P.},
  journal   = {Phys. Rev. Lett.},
  title     = {{Coexistence of BCS- and BEC-Like Pair Structures in Halo Nuclei}},
  year      = {2007},
  month     = {Jul},
  pages     = {022506},
  volume    = {99},
  doi       = {10.1103/PhysRevLett.99.022506},
  issue     = {2},
  numpages  = {4},
  publisher = {American Physical Society},
  url       = {https://link.aps.org/doi/10.1103/PhysRevLett.99.022506},
}

@misc{Supp,
    note = {See Supplemental Material at [URL] for details on the neural-network architecture, training setup, and convergence tests of \textit{FeynmanNet}.}
}

\end{document}